\documentclass{article}
\usepackage{graphicx}
\usepackage{textcomp}
\usepackage{color}
\usepackage{authblk}
\newcommand{\unit}[1]{\ensuremath{\, \mathrm{#1}}}
\newcommand{\unitt}[1]{~{#1}}

\begin{document}
\author[1]{J.E.~van den Berg}
\author[1]{S.C.~Mathavan}
\author[1]{C.~Meinema}
\author[1]{J.~Nauta}
\author[1]{T.H.~Nijbroek}
\author[1]{K.~Jungmann}
\author[2]{H.L.~Bethlem}
\author[1]{S.~Hoekstra}

\affil[1]{University of Groningen, Faculty of Mathematics and Natural 
Sciences, Zernikelaan 25, 9747 AA, Groningen, The Netherlands}
\affil[2]{LaserLaB, Department of Physics and Astronomy, VU University 
Amsterdam, De Boelelaan 1081, 1081 HV, Amsterdam, The Netherlands}

\title{Traveling-wave deceleration of SrF molecules}

\maketitle
\begin{abstract}
We report on the production, deceleration and detection of a SrF molecular beam. The molecules are captured from a supersonic expansion and are decelerated in the X$^2\Sigma^+ (v=0, N=1)$ state. We demonstrate the removal of up to 40\% of the kinetic energy with a 2 meter long modular traveling-wave decelerator. Our results demonstrate a crucial step towards the preparation of ultracold gases of heavy diatomic molecules for precision spectroscopy.
\end{abstract}

\section{Introduction}
\subsection{Motivation}
Heavy diatomic molecules have a high potential as sensitive probes of parity 
violation~\cite{Sushkov:1978vja, 
Kozlov:1991uh, Demille:2008he}. The close proximity of opposite parity states 
offered by selected molecular systems causes a huge enhancement of weak 
interaction effects compared to atoms~\cite{Borschevsky:2012ic}. However, so far 
parity violation at atomic energy scales has only been studied 
in atoms, but not yet in molecules. One of the biggest hurdles is of 
experimental nature: methods have to be developed to control and prepare a 
sufficiently large number of molecules in a well-defined quantum state, as the 
starting point for any sensitive measurement. 

Recently, successful experiments probing fundamental physics using the special 
properties of selected diatomic molecules have been performed using molecular 
beams, for example to put a limit on the possible size of the electric dipole 
moment of the electron~\cite{Hudson:2011hs, Baron:2013eja}. Molecular beams 
offer relatively high density, but limited interaction time. For a next 
generation of precision measurements using heavy diatomic molecules we explore 
the possibilities to make use of a cold gas of trapped molecules. Trapped 
molecules potentially offer very long coherence times that can more than compensate
for the lower number of molecules; such that weak interaction effects can be studied as 
perturbations on a coherently evolving superposition of molecular states.

A diverse range of techniques to decelerate, cool and trap molecules has been 
developed in the past decade. The motivation to create cold molecular gases 
ranges from the possibility to study the strong long-range dipole-dipole 
interactions~\cite{Yan:2013uf}, to molecular collision studies with an 
unprecedented level of control of the internal state of the 
molecules~\cite{2006Sci...313.1617G, 2012Sci...338.1060K}, tests of 
fundamental physics such as the electron EDM as well as the study of possible 
variations of fundamental constants~\cite{Flambaum:2007dw, Bethlem:2009kt}. More 
applications and an overview of the techniques to produce cold molecules have been
addressed in recent review articles~\cite{Meerakker:2012ChemRev,Lemeshko:2013kk}.

The SrF molecule is a sensitive candidate to study weak interaction effects, 
specifically nuclear-spin dependent effects and the anapole moment of the nucleus. 
Coincidentally, the SrF molecule is prototypical of a group of alkaline-earth 
monohalide molecules whose properties allow for a certain amount of laser 
cooling due to favorable Franck-Condon factors. Molecular laser cooling was 
first demonstrated to work for SrF~\cite{2010Natur.467..820S}, and other 
suitable molecules include YO~\cite{Hummon:2013gm}, 
CaF~\cite{Zhelyazkova:2013wb} and probably YbF~\cite{2013NJPh...15e3034T} and RaF~\cite{Isaev:2010gb}. However, 
since the transitions used are not completely closed due to residual leaks to 
excited vibrational levels, it is challenging to provide a sufficiently strong 
force by scattering photons to completely cool and trap the molecules (even from 
a pre-cooled molecular beam).

Due to its sizeable dipole moment of 3.5\unitt{Debye}, the lowest rotational levels of the SrF molecule in its 
$^2\Sigma^+ (v=0)$ absolute ground state have a reasonable Stark shift in external electric fields, and are therefore 
amenable to Stark deceleration. Our approach to create ultracold samples of SrF molecules is to combine a traveling-wave 
Stark decelerator with molecular laser cooling. The decelerator can capture an internally cold sample from a supersonic 
expansion and bring this sample to rest in the laboratory frame. By scattering $\sim1000$ photons the sample can be to 
brought to the Doppler temperature limit. As the molecules remain trapped during the entire deceleration process, 
losses of slow molecules due to transverse beam spreading are absent. 

In this article we describe the creation of a supersonic beam of SrF molecules, 
the traveling-wave guiding and deceleration of these molecules and the detection 
of the decelerated molecules using laser induced fluorescence. We demonstrate 
all key components needed to fully stop and statically trap a pulsed beam of SrF 
molecules. For the measurements presented here a modular decelerator of 2 meter 
length has been employed; for complete stopping this decelerator has to be extended 
to a length of 5 meter. The good agreement of our measurements with the numerical 
simulations that we have performed demonstrates that traveling-wave decelerators 
of this length are indeed completely stable: molecules that fall within its 
phase-space acceptance are guided or decelerated without losses. 

\subsection{Stark deceleration of heavy diatomics}
\begin{figure}[ht!]
 \centering
 \includegraphics[width=\columnwidth]{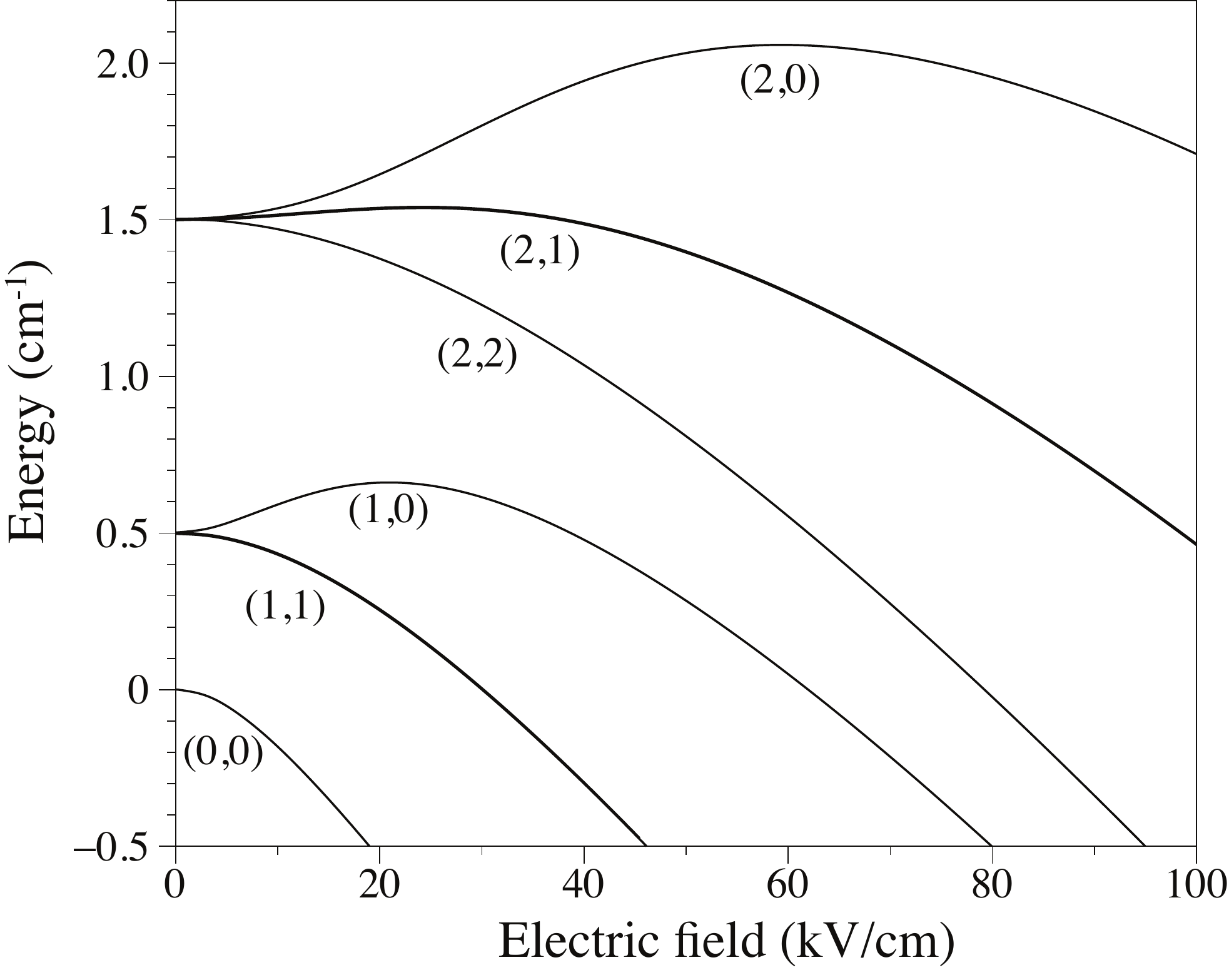}
 \caption{The Stark shift of the three lowest rotational levels of the $^{88}$SrF molecule in its $\mathrm{X}^{2}\Sigma^{+}$ groundstate. The levels are labeled $(N,M_N)$ where $N$ is the rotational number and $M_N$ the projection of $N$ on the electric field axis. Hyperfine structure is not visible on this scale.}
 \label{fig:starkcurve}
\end{figure}

Stark deceleration uses the fact that the energy levels of molecules with an 
electric dipole moment shift in an external electric field. When such a molecule 
travels through an inhomogeneous electric field it loses kinetic energy while 
it gains potential energy from climbing up the electric field hill. If this 
process of climbing the potential hill is repeated many times, the kinetic 
energy loss can be considerable. In practice this is done for light molecules in 
a traditional Stark decelerator by repeatedly switching the polarity of an array of 
high-voltage electrodes through which the molecular beam traverses. In 
comparison to lighter molecules with a linear Stark shift, heavy diatomics such 
as SrF are more difficult to decelerate. First, for a fixed initial velocity and 
deceleration strength, the large mass requires a rather long decelerator. This 
puts high demands on the stability of the deceleration 
process~\cite{Meerakker:2006tx}. Second, related to the small spacing of 
rotational levels, at modest electric 
fields (20-30 kV/cm) the slope of the Stark curve of the lowest rotational 
levels turns negative; at higher fields these states are all so-called 
high-field seekers, and are attracted to the high field at the electrodes. The 
deceleration of molecules in high-field seeking states has proven to be very 
challenging~\cite{Bethlem:2006vx}. At the turning point the maximum Stark shift  
for the lowest rotational states is only a fraction of a wavenumber. This is 
illustrated in Figure~\ref{fig:starkcurve} for the lowest rotational states of SrF 
in its $X^2\Sigma^+(v = 0)$ electronic and vibrational ground state. The 
unfavorable Stark shift and the high mass result in highly inefficient  
deceleration of SrF with a traditional Stark decelerator. Therefore we have built a 
traveling-wave decelerator~\cite{2010PhRvA..81e1401O} consisting of a sequence of ring 
electrodes which are used to propagate 3D electric traps that can hold bunches of 
molecules, thereby mitigating most of the losses that traditional Stark 
decelerators have. Motivated by similar arguments, an experiment has been performed on the deceleration 
of YbF molecules in a 0.5 meter traveling-wave decelerator~\cite{2012PhRvA..86b1404B}. Recently the static trapping of ammonia molecules in a traveling-wave decelerator has been demonstrated~\cite{QuinteroPerez:2013bk,2013PhRvA..88d3424J}.

The working principle of the traveling-wave decelerator has been discussed
 in literature~\cite{2010PhRvA..81e1401O,Meek:2011bpa,2012EPJD...66..235V}. Here we 
briefly summarize the essential features. 
A sequence of ring-shaped electrodes is supplied with a sinusoidal voltage that varies in space and time such that there 
is a phase shift of 45 degrees between consecutive electrodes. Electric field minima and maxima are 
created along the axis of the ring structure as shown in Figure~\ref{fig:experiment}. Inside the periodic structure, 
the electric field minima will appear at regular intervals. These minima 
can be moved along the decelerator axis by applying these oscillating voltages to the rings in 
such a way that a traveling wave is created. Molecules in low-field-seeking 
states will experience these minima as three-dimensional traps. The speed of the traps 
can be changed by varying the frequency of the oscillating voltage.

Here we report first time deceleration of SrF molecules. We show that the modular 
traveling-wave decelerator of 2 meter length (4 modules of 0.5 meters each) is stable in operation and construction. This is of crucial importance for the extension of the device to $\sim 5$ meter length which is required to completely stop SrF molecules.

\section{Experimental setup}
\begin{figure*}
 \centering
 \includegraphics[width=\textwidth]{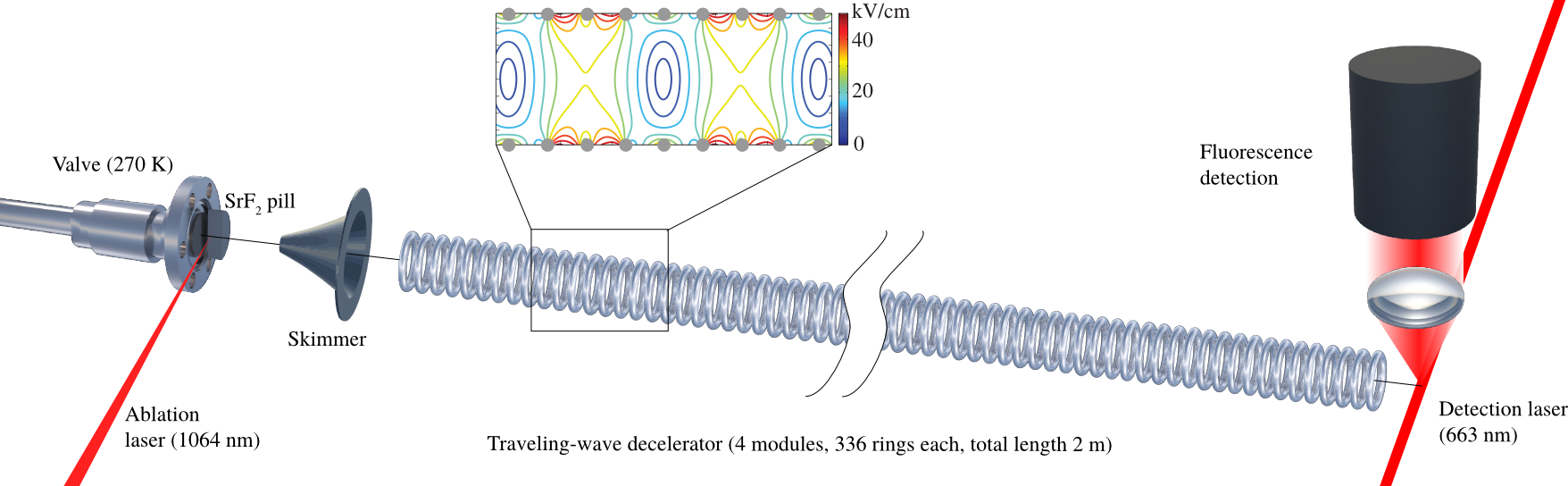}
 \caption{A schematic view of the experiment. SrF molecules are created by laser ablation from a pill. A pulsed supersonic Xe expansion cools the SrF molecules and takes them to the ring structure where the molecules are guided or decelerated. Further downstream the SrF is detected by laser induced fluorescence.}
 \label{fig:experiment}
\end{figure*}

Figure~\ref{fig:experiment} shows a schematic view of our deceleration 
experiment consisting of a supersonic expansion source, the decelerator and a 
laser-induced fluorescence detection zone. A 100\unitt{mJ} pulse 
 from a Nd:YAG laser at its 
principal wavelength is used to ablate SrF radicals from a home-pressed pill 
consisting of 90\% $\rm{SrF_2}$ and 10\% B into a supersonic expansion of Xe gas 
from a pulsed General Valve. The addition of boron is found to make the pill 
much more stable~\cite{2009RScI...80k3111T}. The valve is operated at 10 Hz and it 
is cooled with cold nitrogen gas to -30\unitt{\textcelsius}, resulting in an 
average molecular beam speed of 300\unitt{m/s}. From a rotational spectrum we find a rotational 
temperature of $\sim$~10\unitt{K}. Upon optimisation of the source we expect to be able 
to reach a comparable output as a supersonic source for YbF 
molecules~\cite{Tarbutt:2002wv}. The molecular 
beam enters the decelerator located 125\unitt{mm} downstream of the valve through a 
skimmer with a diameter of $2 \unit{mm}$, placed 60\unit{mm} from the ablation 
spot.

The dimensions of the ring-shaped electrodes and the method in which these are 
positioned relative to each other is based on the design of Meek \textit{et 
al.}~\cite{Meek:2011bpa}. Since we require a rather long decelerator however, we 
combine four modules of 50\unitt{cm} in order to form a horizontally oriented 
decelerator with a length of 2.016\unitt{m}. Each module contains 336 
ring-shaped electrodes made of tantalum wire of 0.6\unitt{mm} diameter. The 
electrodes are mounted on eight 8\unitt{mm} diameter cylindrical stainless steel rods which are 
placed in an octagonal pattern on the outside of a circle with circumference of 
26\unitt{mm}. The gap between two 
consecutive rings is 0.9\unitt{mm}, resulting in a periodicity of 12 \unit{mm}. 

\subsection{High-voltage electronics}
Arbitrary waveforms are generated using DACQ8150 Acquitek PCI cards and 
amplified by eight custom high-voltage amplifiers from Trek Inc. 
These amplifiers have an output voltage of maximally \textpm 5\unitt{kV} at frequencies 
between 30\unitt{kHz} and DC and can handle the full capacitance of a 5 meter 
long decelerator. This makes it possible to decelerate molecules to a complete stop and 
then keep the molecules in a static trap. Furthermore, the arbitrary waveforms 
can be used to tailor the shape of the potential at will or to manipulate the 
trap as we like~\cite{2013PhRvA..88d3424J}.

\subsection{Detection}
After deceleration and 116.5\unitt{mm} of free flight, the molecules are 
state-selectively detected  using a resonant laser-induced fluorescence scheme. 
Light at 663.3\unitt{nm} to drive the $A ^2\Pi_{1/2} (v=0) \leftarrow X 
^2\Sigma^+ (v=0) \; P(1/2),Q(1/2)$ transitions, which probes the population in the lowest low-field seeking $N=1$ rotational level, is generated using an external-cavity diode laser. It is this 
state which is the starting point for a parity violation measurement, and the 
state for which laser cooling has been demonstrated to work. The $N=1$ state is also 
the best choice for deceleration given the electric field strength available to us. 
Approximately 10\unitt{mW} of laser light is used to lock the laser to  
absorption line R(115) (6-6) a9 in molecular iodine  enclosed in a quartz cell heated to 
300\unitt{\textcelsius}. We identify this line using frequency-modulated 
doppler-free saturated-absorption spectroscopy. The light used for SrF detection is sent through an EOM 
driven at 41\unitt{MHz} for the creation of sidebands that overlap with the 
resolved hyperfine levels in the SrF X$^2\Sigma^+ (v=0,N=1)$ state~\cite{2010Natur.467..820S}. It 
reaches the detection chamber through a single mode optical fiber where it is 
directed through the detection region using two mirrors. The beam is 
retro-reflected such that it crosses the molecular beam at right angles twice. Typically, the laser power is 
2\unitt{mW} and the beam diameter is 3\unitt{mm}. A system of lenses focusses 
the fluorescent light through an interference filter onto the cathode of a 
photomultiplier tube (PMT). We have an overall detection efficiency of 2\%. The 
arrival time of the PMT pulses is recorded by a multi-channel analyzer with $1\unit{\mu s}$ bins. 
The time-of-flight spectra are started with the ablation pulse.

\section{Results}\label{results}
Here we present the results obtained from guiding and decelerating SrF(1,0) molecules with the traveling wave decelerator.
\begin{figure*}
 \centering
 \includegraphics[width=\textwidth]{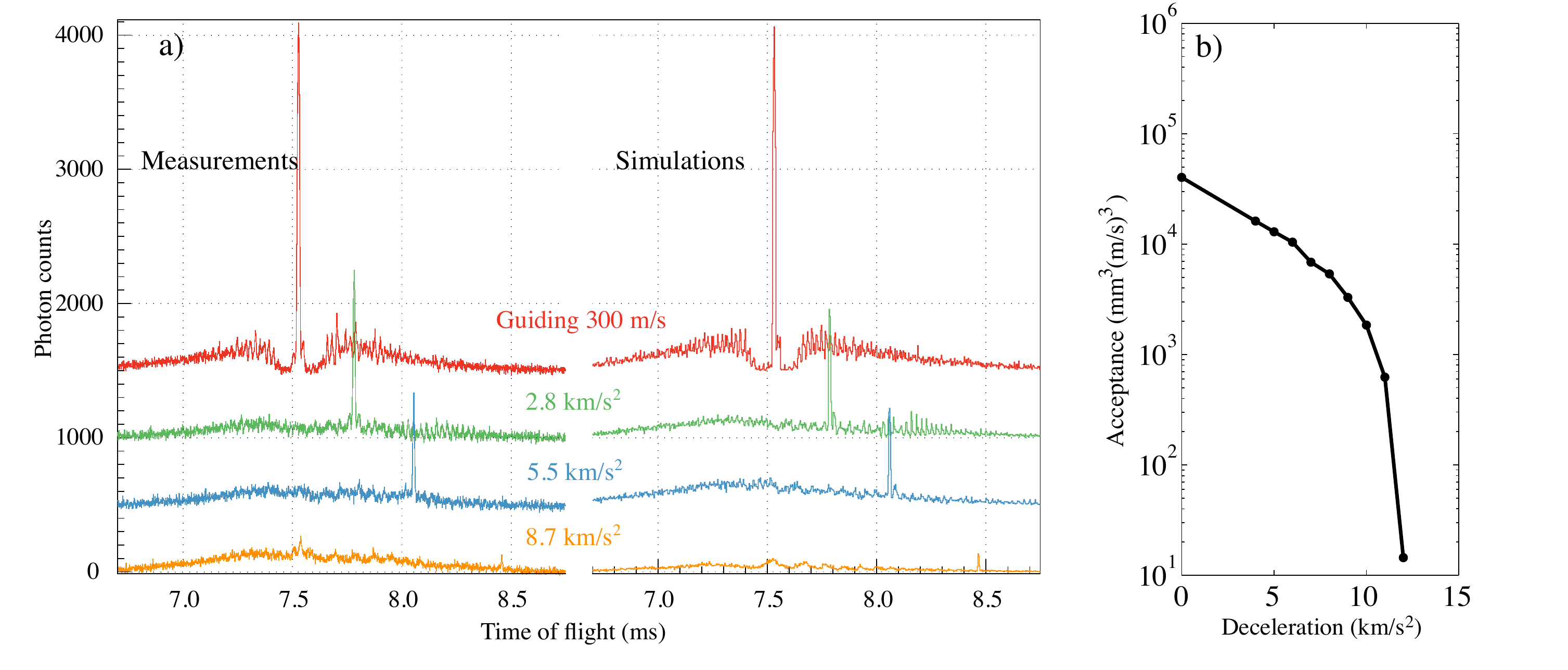}
 \caption{a) Measured (left) and simulated (right) time-of-flight profiles for 
SrF (1,0) molecules after 2\unitt{m} of deceleration. The different measurements 
have been given a vertical offset for clarity. b) The 3D acceptance of the 
traveling-wave decelerator as a function of deceleration strength, as found from 
numerical trajectory calculations~\cite{2012EPJD...66..235V}.}
 \label{fig:decel300}
\end{figure*}

The left side of Figure~\ref{fig:decel300}a shows measured time-of-flight (TOF) spectra 
of SrF(1,0) molecules. For each spectrum a small constant background has been subtracted. The spectra have been given a vertical offset for clarity. The mean speed of the beam was 300\unitt{m/s} and the starting speed for the decelerator waveforms was set accordingly. The arrival time of the molecules is plotted for guiding at 
300\unitt{m/s} and a range of deceleration strengths, each of which was 
constant throughout the full 2\unitt{m} length of the decelerator. The delayed 
arrival of the main package of molecules with increasing deceleration strength 
can clearly be seen. At the highest deceleration of 8.7\unit{km/s^2} the package 
of decelerated molecules is arriving around 8.45\unitt{ms}, corresponding to 
molecules that were decelerated from 300\unitt{m/s} to 234\unitt{m/s}.  This 
corresponds to a removal of 40\% of the kinetic energy.

The TOF profiles show very distinct features which are best seen in the guiding 
measurement (Figure~\ref{fig:decel300}a upper trace). At 7.5\unitt{ms} there is a sharp peak that corresponds to molecules that are trapped in a moving electric trap. 
The small maxima next to the main peak are due to molecules that were trapped in potential wells adjacent to the central well.
 Because of the small size of 
the SrF cloud at the entrance of the decelerator almost all SrF molecules are 
loaded into a single potential bucket. The molecules that are collected in these 
wells leave behind a big dimple in the final TOF spectrum which is broadened 
because of flight time. The molecules that fall outside the longitudinal 
phase-space acceptance but remain inside the transverse phase-space acceptance of the 
decelerator are not captured in the potential well, but are still guided through 
and detected. These molecules end up in the broad wings that are observable in 
the guiding signal around 7.3 and 7.8 \unitt{ms}.  The modulation that is 
observable on these broad wings is caused by bunching in phase space on the 
moving potential hills in between the electric field minima.

We have calculated~\cite{2012EPJD...66..235V} the phase space acceptance 
of the decelerator for different deceleration strengths as shown in Figure~\ref{fig:decel300}b. 
As an addition to this we have now simulated the TOF spectra as well, including the effects of the 
shape of the initial distribution. By comparing the simulations to the measurements, 
we derive the properties of the molecular beam after the supersonic expansion. We find a longitudinal velocity 
distribution with a mean speed of 300\unitt{m/s} and a standard deviation of 20\unitt{m/s}. Furthermore the 
starting size of the molecular packet is of order 1\unitt{mm}, which 
matches to the spot size of the ablation laser on the pill. As displayed on the right side of 
Figure~\ref{fig:decel300}a all the features of the 
measurements are reproduced in the simulations. The smaller peaks for higher deceleration strengths 
can be understood from the decreasing phase space acceptance and are not due to losses in the decelerator.
 The simulations have been scaled 
on the guiding signal to match the intensity of the non-decelerated part of the 
detected molecules. The height of the decelerated peak in the simulation for the 
strongest deceleration is higher than measured, which we attribute to suboptimal 
experimental conditions. During the experiments the laser ablation power 
gradually reduced and the poor regulation of the valve temperature caused the supersonic expansion 
to change slightly. The experiment with the strongest deceleration is the most sensitive 
to these effects, because of the increased sensitivity to the timing of the 
molecular pulse relative to the start of the deceleration, since the phase-space acceptance is rather small. As a consequence of these measurements we are implementing an improvement of the temperature 
regulation of the cooled supersonic expansion.

Each spectrum shown is the sum of 10240 ablation shots, except for the strongest 
deceleration strength which is the average of twice as many shots because of frequent
unlocking of the spectroscopy laser. In a single 
shot only a few photons are detected, indicative of a 
rather low beam density. Being cautious not to damage to the ablation pill, we 
have used the lowest possible ablation power that enabled us to still measure with a 
good signal-to-background ratio. Obtaining the time-of-flight traces in 
Figure~\ref{fig:decel300} took about 20 minutes per trace. In later 
experiments we have seen that the ablation power could be increased significantly 
without damaging the pill, giving a correspondingly higher yield of 
SrF molecules. We are currently optimizing the ablation geometry to provide better 
rotational cooling of the molecular beam, thereby increasing the fraction of molecules 
in the (1,0) state. 

\section{Conclusions}
Here we report the first Stark deceleration of SrF molecules. These 
measurements demonstrate the successful combined operation of the supersonic expansion 
source of SrF molecules, the 2\unitt{m} long modulator decelerator and the laser-induced 
fluorescence detection. Extension of the decelerator to 5\unitt{m} is underway, 
and it is expected to deliver completely stopped bunches of SrF molecules. Such a 
decelerator is a general device, extending the range of molecules that can be 
decelerated and trapped. These include relatively heavy molecules with an unfavorable Stark shift such as SrF, YbF 
and PbO, larger (bio)molecules such as benzonitrile~\cite{Filsinger2009}, but also lighter molecules that have so far 
eluded Stark deceleration such as water. We expect 
these cold samples of molecules, especially in combination with further cooling 
methods, to be the starting point for many exciting future experiments.

\section{Acknowledgments}
We acknowledge the expert technical assistance of Leo Huisman, Imko Smid and the KVI mechanical workshop. 
This work is part of the research programme of the Foundation for Fundamental 
Research on Matter (FOM), which is part of the Netherlands Organization for 
Scientific Research (NWO) (FOM Programs nr. 114 and 125, Projectruimte 11PR2858, VIDI 680-47-519). 
\bibliographystyle{elsarticle-num} 

\begin{thebibliography}{10}
\expandafter\ifx\csname url\endcsname\relax
  \def\url#1{\texttt{#1}}\fi
\expandafter\ifx\csname urlprefix\endcsname\relax\def\urlprefix{URL }\fi
\expandafter\ifx\csname href\endcsname\relax
  \def\href#1#2{#2} \def\path#1{#1}\fi

\bibitem{Sushkov:1978vja}
O.~P. Sushkov, V.~V. Flambaum, {Parity breaking effects in diatomic molecules},
  Sov. Phys. JETP 48 (1978) 608.

\bibitem{Kozlov:1991uh}
M.~G. Kozlov, L.~N. Labzovskii, A.~O. Mitrushchenkov, {Parity nonconservation
  in diatomic molecules in a strong constant magnetic field}, Sov. Phys. JETP
  73 (1991) 415.

\bibitem{Demille:2008he}
D.~DeMille, S.~B. Cahn, D.~Murphree, D.~A. Rahmlow, M.~G. Kozlov, {Using
  Molecules to Measure Nuclear Spin-Dependent Parity Violation}, Physical
  Review Letters 100~(2) (2008) 023003.

\bibitem{Borschevsky:2012ic}
A.~Borschevsky, M.~Ilias, V.~A. Dzuba, K.~Beloy, V.~V. Flambaum,
  P.~Schwerdtfeger, {Nuclear-spin-dependent parity violation in diatomic
  molecular ions}, Physical Review A 86~(5) (2012) 050501(R).

\bibitem{Hudson:2011hs}
J.~J. Hudson, D.~M. Kara, I.~J. Smallman, B.~E. Sauer, M.~R. Tarbutt, E.~A.
  Hinds, {Improved measurement of the shape of the electron}, Nature 473~(7348)
  (2011) 493--496.

\bibitem{Baron:2013eja}
{The ACME Collaboration}, {Order of Magnitude Smaller Limit on the Electric
  Dipole Moment of the Electron}, Science 343 (2014) 269.

\bibitem{Yan:2013uf}
B.~Yan, S.~A. Moses, B.~Gadway, J.~P. Covey, K.~R.~A. Hazzard, A.~M. Rey, D.~S.
  Jin, J.~Ye, {Realizing a lattice spin model with polar molecules}, Nature
  501~(7468) (2013) 521--525.

\bibitem{2006Sci...313.1617G}
J.~J. Gilijamse, S.~Hoekstra, S.~Y.~T. van~de Meerakker, G.~C. Groenenboom,
  G.~Meijer, {Near-Threshold Inelastic Collisions Using Molecular Beams with a
  Tunable Velocity}, Science 313~(5) (2006) 1617--1620.

\bibitem{2012Sci...338.1060K}
M.~Kirste, X.~Wang, H.~C. Schewe, G.~Meijer, K.~Liu, A.~van~der Avoird,
  L.~M.~C. Janssen, K.~B. Gubbels, G.~C. Groenenboom, S.~Y.~T. van~de
  Meerakker, {Quantum-State Resolved Bimolecular Collisions of
  Velocity-Controlled OH with NO Radicals}, Science 338~(6) (2012) 1060.

\bibitem{Flambaum:2007dw}
V.~V. Flambaum, M.~Kozlov, {Enhanced sensitivity to the time variation of the
  fine-structure constant and m(p)/m(e) in diatomic molecules}, Physical Review
  Letters 99 (2007) 150801.

\bibitem{Bethlem:2009kt}
H.~L. Bethlem, W.~Ubachs, {Testing the time-invariance of fundamental constants
  using microwave spectroscopy on cold diatomic radicals}, Faraday Discussions
  142 (2009) 25.

\bibitem{Meerakker:2012ChemRev}
S.~Y.~T. van~de Meerakker, H.~L. Bethlem, N.~Vanhaecke, G.~Meijer,
  {Manipulation and control of molecular beams}, Chemical Reviews 112 (2012)
  4828--4878.

\bibitem{Lemeshko:2013kk}
M.~Lemeshko, R.~V. Krems, J.~M. Doyle, S.~Kais, {Manipulation of Molecules with
  Electromagnetic Fields}, Molecular Physics 111 (2013) 1648--1682.

\bibitem{2010Natur.467..820S}
E.~S. Shuman, J.~F. Barry, D.~DeMille, {Laser cooling of a diatomic molecule},
  Nature 467~(7) (2010) 820--823.

\bibitem{Hummon:2013gm}
M.~T. Hummon, M.~Yeo, B.~K. Stuhl, A.~L. Collopy, Y.~Xia, J.~Ye, {2D
  Magneto-Optical Trapping of Diatomic Molecules}, Physical Review Letters
  110~(14) (2013) 143001.

\bibitem{Zhelyazkova:2013wb}
V.~Zhelyazkova, A.~Cournol, T.~E. Wall, A.~Matsushima, J.~J. Hudson, E.~A.
  Hinds, M.~R. Tarbutt, B.~E. Sauer, {Laser cooling and slowing of CaF
  molecules}, arXiv.org\href {http://arxiv.org/abs/1308.0421v1}
  {\path{arXiv:1308.0421v1}}.

\bibitem{2013NJPh...15e3034T}
M.~R. Tarbutt, B.~E. Sauer, J.~J. Hudson, E.~A. Hinds, {Design for a fountain
  of YbF molecules to measure the electron's electric dipole moment}, New
  Journal of Physics 15~(5) (2013) 3034.

\bibitem{Isaev:2010gb}
T.~A. Isaev, S.~Hoekstra, R.~Berger, {Laser-cooled RaF as a promising candidate
  to measure molecular parity violation}, Physical Review A 82~(5) (2010)
  052521.

\bibitem{Meerakker:2006tx}
S.~Y. T. v.~d. Meerakker, N.~Vanhaecke, H.~L. Bethlem, G.~Meijer, {Transverse
  stability in a Stark decelerator}, Physical Review A 73 (2006) 023401.

\bibitem{Bethlem:2006vx}
H.~L. Bethlem, M.~R. Tarbutt, J.~K. Upper, D.~Carty, K.~Wohlfart, E.~A. Hinds,
  G.~Meijer, {Alternating gradient focusing and deceleration of polar
  molecules}, Journal Of Physics B-Atomic Molecular And Optical Physics 39~(16)
  (2006) R263--R291.

\bibitem{2010PhRvA..81e1401O}
A.~Osterwalder, S.~A. Meek, G.~Hammer, H.~Haak, G.~Meijer, {Deceleration of
  neutral molecules in macroscopic traveling traps}, Physical Review A 81~(5)
  (2010) 051401(R).

\bibitem{2012PhRvA..86b1404B}
N.~E. Bulleid, R.~J. Hendricks, E.~A. Hinds, S.~A. Meek, G.~Meijer,
  A.~Osterwalder, M.~R. Tarbutt, {Traveling-wave deceleration of heavy polar
  molecules in low-field-seeking states}, Physical Review A 86~(2) (2012)
  21404.

\bibitem{QuinteroPerez:2013bk}
M.~Quintero-P{\'e}rez, P.~Jansen, T.~Wall, J.~E. van~den Berg, S.~Hoekstra,
  H.~L. Bethlem, {Static Trapping of Polar Molecules in a Traveling Wave
  Decelerator}, Physical Review Letters 110~(13) (2013) 133003.

\bibitem{2013PhRvA..88d3424J}
P.~Jansen, M.~Quintero-P{\'e}rez, T.~E. Wall, J.~E. van~den Berg, S.~Hoekstra,
  H.~L. Bethlem, {Deceleration and trapping of ammonia molecules in a
  traveling-wave decelerator}, Physical Review A 88~(4) (2013) 43424.

\bibitem{Meek:2011bpa}
S.~A. Meek, M.~F. Parsons, G.~Heyne, V.~Platschkowski, H.~Haak, G.~Meijer,
  A.~Osterwalder, {A traveling wave decelerator for neutral polar molecules},
  Review Of Scientific Instruments 82~(9) (2011) 093108--093108--9.

\bibitem{2012EPJD...66..235V}
J.~E. van~den Berg, S.~H. Turkesteen, E.~B. Prinsen, S.~Hoekstra, {Deceleration
  and trapping of heavy diatomic molecules using a ring-decelerator}, The
  European Physical Journal D 66~(9) (2012) 235.

\bibitem{2009RScI...80k3111T}
M.-F. Tu, J.-J. Ho, C.-C. Hsieh, Y.-C. Chen, {Intense SrF radical beam for
  molecular cooling experiments}, Review Of Scientific Instruments 80~(1)
  (2009) 3111.

\bibitem{Tarbutt:2002wv}
M.~R. Tarbutt, J.~J. Hudson, B.~E. Sauer, E.~A. Hinds, V.~A. Ryzhov, V.~L.
  Ryabov, V.~F. Ezhov, {A jet beam source of cold YbF radicals}, Journal Of
  Physics B-Atomic Molecular And Optical Physics 35 (2002) 5013.

\bibitem{Filsinger2009}
F.~Filsinger, J.~K\"upper, G.~Meijer, L.~Holmegaard, J.~H.~Nielsen, I.~Nevo, J.~L.~Hansen, H.~Stapelfeldt, 
 {Quantum-state selection, alignment, and orientation of large molecules using static electric and laser fields}, The 
 Journal of Chemical Physics 131~(6) (2009)

\end{thebibliography}

\end{document}